\documentclass{pnastwo}
\usepackage{graphicx}
\usepackage{amssymb,amsfonts,amsmath,gensymb}
\usepackage{color}

\newcommand{\updated}[1]{{\color{blue} #1 }}

\begin{document}

\title{Identifying the mechanism for superdiffusivity in mouse fibroblast motility} 

\author{G Passucci\affil{1}{Physics Department, Syracuse University, Syracuse, NY 13244},
ME Brasch\affil{2}{Biomedical and Chemical Engineering Department, Syracuse University, Syracuse, NY 13244}, 
JH Henderson\affil{2}{}\affil{3}{Syracuse Biomaterials Institute, Syracuse University, Syracuse, NY 13244},
V Zaburdaev\affil{4}{Institute of Supercomputing Technologies, Lobachevsky State University of Nizhny Novgorod, 603140 Nizhny Novgorod, Russia},
\and
ML Manning\affil{1}{}\affil{3}{}}

\contributor{Submitted to Proceedings of the National Academy of Sciences
of the United States of America}

%---------------------------------------------------------------------------------------

\maketitle % The \maketitle command is necessary to build the title   page

\begin{article}

\begin{abstract}
 We seek to characterize the motility of mouse fibroblasts on 2D substrates. Utilizing automated tracking techniques, we find that cell trajectories are super-diffusive, where displacements scale faster than $t^{1/2}$ in all directions. Two mechanisms have been proposed to explain such statistics in other cell types: run and tumble behavior with L\'{e}vy-distributed run times, and ensembles of cells with heterogeneous speed and rotational noise. We develop an automated toolkit that directly compares cell trajectories to the predictions of each model and demonstrate that ensemble-averaged quantities such as the mean-squared displacements and velocity autocorrelation functions are equally well-fit by either model. However, neither model correctly captures the short-timescale behavior quantified by the displacement probability distribution or the turning angle distribution. We develop a hybrid model that includes both run and tumble behavior and heterogeneous noise during the runs, which correctly matches the short-timescale behaviors and indicates that the run times are not L\'{e}vy distributed. The analysis tools developed here should be broadly useful for distinguishing between mechanisms for superdiffusivity in other cells types and environments.
\end{abstract}

%\keywords{monolayer | structure | x-ray reflectivity | molecular electronics}

\section{Introduction}

\dropcap{C}ell motility is an integral part of biological processes such as morphogenesis \cite{Friedl2004}, wound healing \cite{Bindschadler2007} , and cancer invasion \cite{Kraning-Rush2013}. But what are the rules that govern how cells move? Cell migration involves a multitude of organelles and signaling pathways \cite{Lauffenburger1996} and therefore a fruitful, bottom-up approach studies correlations between cell motion and sub-cellular processes that govern motility, including surface interactions \cite{Zaman2006}, integrin signaling pathways \cite{Yamada1995}, or formation of focal adhesions \cite{Schneider2008}. 

An alternate approach with recent successes is to develop simple models at the cellular scale that can help identify a coarse-grained set of rules that govern cell migration in specific cell types.  One such class of models, composed of self-propelled (SPP) or active Brownian particles ~\cite{Vicsek1995} has been used to make predictions about the motion of biological cells in many contexts, including density fluctuations \cite{Zhang2010}, formation of bacterial colonies \cite{Czirok1996}, \updated{and} both confined~\cite{Henkes2011}, and expanding monolayers~\cite{Poujade2007}. 

These SPP models represent each cell as a particle that moves by generating active force on a substrate, which acts along a specified vector $\hat{\theta}$. Therefore, the parameters for the model specify both the magnitude of the force as well as how the orientation of the force changes with time. Given the ubiquity and usefulness of these models, one would like to have a standard framework for extracting these parameters from experimental data for all trajectories. In the past this has often been accomplished by analyzing ensemble-averaged features of cell trajectories.

One such quantity is the time averaged mean-squared displacement (MSD), which is the squared displacement between positions $\vec{r}(t)$ and $\vec{r}(t + dt)$ averaged over all starting times $t$ and the ensemble of trajectories. This yields the MSD as a function of timescale, $\langle(r(t+dt))-r(t))^2\rangle \propto dt^{\alpha}$. Ballistic motion, which corresponds to a cell moving in a straight line at constant speed, corresponds to $\alpha = 2$. Diffusive motion, where a cell executes a random walk with no time correlation in orientation, corresponds to $\alpha = 1$. In non-active matter at low densities, thermal fluctuations generically induce diffusive behavior at long timescales. In contrast, many cell types, including T-cells \cite{Harris2012}, Hydra cells \cite{Upadhyaya2000}, breast carcinoma cells \cite{Metzner2015}, and swarming bacteria \cite{Ariel} display super-diffusive dynamics, defined as trajectories with a MSD exponent between $1 < \alpha < 2$. 

Several authors have proposed explanations for why super-diffusive migration might be beneficial in biological systems. For example, super-diffusive trajectories are well known for being the optimal search strategy for randomly placed sparse targets \cite{Raposo2009,Viswanathan1999}, and have been found in animal foraging and migration patterns in jellyfish \cite{Reynolds2014}, albatross, and bumblebees \cite{Edwards2007}. In the context of cell biology, superdiffusive migration implies that cells are covering new areas more quickly than they would if they were executing a simple random walk. 

Although super-diffusive dynamics are commonly observed in \textit{in vitro} experiments, the fundamental mechanism that generates anomalous diffusion in cell trajectories remains unclear. Pinpointing the mechanism would allow biology researchers to better isolate the signaling pathways that govern these processes. 

Although one might think that simply including the effects of persistent active forces generated by cells would change the long-time behavior, it turns out that standard self-propelled particle models exhibit a fairly sharp crossover from ballistic to diffusive motion, with no extended superdiffusive regime. Since SPP models are commonly used to model cells and superdiffusive dynamics are commonly observed in experiments, we would like to identify the mechanism generating superdiffusitivity to improve the ability of these models to capture cellular phenomena.

%When cell-cell (or particle-particle) interactions are included, standard SPP models generate interesting emergent collective phenomena, such as giant number fluctuations and phase separation.  Moreover, these emergent behaviors are biologically relevant, for example predicting aggregation phenomena in microorganisms \cite{Tailleur2015}.

%SPPs have had extraordinary success in accurately modeling motility induced phase separation (MIPS) exhibited by both synthetic colloids as well as a multitude of microorganisms [Tailleur2015]. 
Standard SPP models include smoothly varying persistent random walkers and standard run-and-tumble particles (RTP) ~\cite{Marchetti2013}. Persistent random walkers obey the following equations of motion for the cell center of mass $r_i$ and the orientation angle $\theta_i$: 

\begin{equation}
\label{spp_v}
\partial_t \vec{r}_i = v_0 \hat{\theta}_i,
\end{equation}
\begin{equation}
\label{spp_t}
\partial_t \theta_i = \eta(t),
\end{equation}
where $\eta(t)$ is a Gaussian white noise ($\langle\eta(t)\eta(t')\rangle = 2D_r \delta(t-t')$). In a standard persistent random walk, the speed $v_0$ and the rotational diffusion coefficient $D_r$, which controls the strength of fluctuations in orientation, are constant. In a standard run-and-tumble model, particles are ballistic during runs, $\partial_t \theta_i = 0$, followed by tumbling events where large changes in orientation occur. Variations of run-and-tumble models are characterized by the distribution of times particles remain in the run state.

Two different classes of modifications to SPP models have been highlighted as being able to generate super-diffusive behavior on long timescales.  The first modification is a heterogeneous speed model, which draws rotational diffusion coefficients and particle speeds from distributions \cite{Metzner2015,Zaburdaev2008}. While persistent random walk models transition from ballistic to diffusive behavior at one characteristic timescale, heterogeneous speed models possess a heterogeneous distribution of crossover timescales, which generates an MSD with a broad superdiffusive regime, though the system becomes diffusive on timescales longer than $1/D_r^{min}$. 

%We implement this type of SPP model with rotational diffusion coefficients and particle speeds drawn from a bivariate Gaussian distribution and compare ensemble averaged trajectory statistics with mouse fibroblast cells as well as the other SPP model which generates superdiffusion, L\'{e}vy walks. 

The second modification is a L\'{e}vy walk model, which is a run-and-tumble model where particles have power law distributed run times:
\begin{equation}
\label{P_tau}
P(\tau) = \frac{\mu}{\tau_o(1+\tau/\tau_o)^{1+\mu}},
\end{equation}
\begin{equation}
\label{tau}
\left<\tau\right> = \frac{\tau_o}{\mu-1},
\end{equation}
with $P(\tau)$ the distribution of run times with mean $<\tau>$ for $ \mu >  1$. \cite{Vasily}. In contrast to the heterogeneous SPP model, super-diffusivity generated by L\'{e}vy walks is not transient, so that the long-time MSD scaling exponent is constant: $MSD \propto dt^{3-\mu}$. 

So which of these models is the ``right" one for a given cell type? By analyzing ensemble-averaged statistics such as the MSD and the velocity autocorrelation function (VACF), one group of researchers was able to show that heterogeneous motility models matched data from breast cancer carcinoma cells \cite{Metzner2015}.  Another group evaluated a different ensemble-averaged quantity, the probability displacement distribution, and used that data to suggest that T-cells were undergoing Levy walks \cite{Harris2012}. We would like to better understand whether these ensemble-averaged quantities are in fact a unique identifier of the underlying mechanism for superdiffusivity. Moreover, we also seek to develop a systematic procedure for using experimental data to constrain both the appropriate mechanism and the optimal model parameters for a specific subtype. To this end, we use automated tracking software to analyze over 1000 mouse fibroblast trajectories. We demonstrate that some ensemble-averaged statistics, such as the MSD and VACF, can not distinguish between mechanisms for superdiffusivity. 

In order to better distinguish, we begin with a very general model for cell dynamics. Although standard SPP models have only two fit parameters, average cell speed $v_0$ and average rotational noise $D_r$, in principal a generalized SPP model could have arbitrary distributions for cell speed $P(v_0)$ and rotational diffusion $P(D_r)$ with arbitrary correlations between them. The heterogeneity motility model from \cite{Metzner2015} is the limit of such a model with Gaussian-distributed $P(v_0)$ and $P(D_r)$, while a L\'{e}vy walk is the limit with a constant $v_0$ and a specialized bimodal $P(D_r)$. Because this is such a large parameter space, we first constrain the functional form of these distributions using specific features of single cell trajectory statistics. We find that the mouse fibroblast data are consistent with run-and-tumble dynamics but the run times are not power-law distributed, confirming that in mouse fibroblasts the mechanism for superdiffusivity is heterogeneous dynamics and not L\'{e}vy walk statistics. The toolkit we have developed here should be useful for pinpointing the origin of superdiffusivity in many other cell types.

\section{Methods}
\section{Mouse fibroblast cell culture}

Cell motility data was collected from C3H10T1/2 mouse fibroblast cells (ATCC) cultured on a flat, gold-coated polymer substrate, prepared as previously described \cite{Yang2013}. Cell nuclei were labled with Hoechst dye and cell motility imaged by time-lapse microscopy under two different temperature conditions, 4 hrs (48 frames) at 30$^{\circ}$C and then 20 hrs (240 frames) at 37$^{\circ}$C (Supplemental Method 1). The resultant motility image stacks were analyzed using the ACT\textit{IV}E image analysis package to track nuclei centers-of-mass~\cite{Baker2014}.

\begin{figure}
\centering
\includegraphics[width=0.5\textwidth]{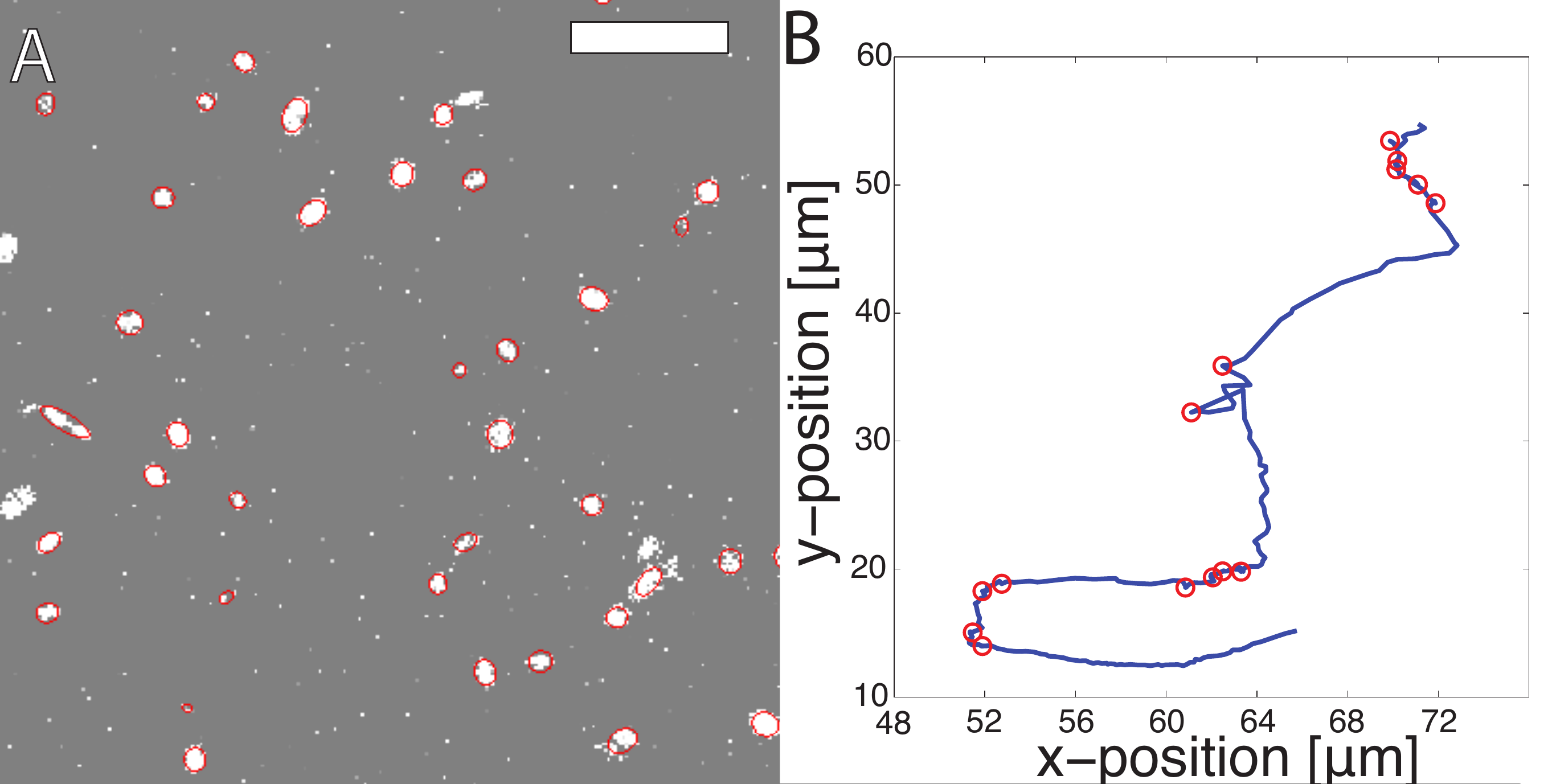}
\caption{\label{mf_example}(A) An example image of nuclei stained with Hoescht dye. The scale bar is 500 $\mu m$. These images were processed using the ACT\textit{IV}E image analysis package to track nuclei centers-of-mass~\cite{Baker2014}, with overlaid best-fit ellipses. (B) A typical cell trajectory with tumbling events indicated by red circles, as identified by the 2D Canny edge detection algorithm.}
\end{figure}

\section{Cell trajectory analysis and particle simulations}

Cell motility was characterized using statistical analysis of cell nuclei trajectories, including MSD, VACF and displacement probability distributions. Tumbling events were identified with a Canny edge detection algorithm. Additional details on cell trajectory analysis can be found in Supplemental Method 2.

\section{Active particle simulations}
This manuscript focuses on two different models for non-interacting active particles. The first model is a L\'{e}vy walk with constant particle speed $v_0$ at all timesteps. Particles execute ballistic runs with zero rotational noise for times $\tau$ drawn from the distribution in Eq. 3 and a mean run time $\left<\tau\right>$ given by Eq. 4. 

The generalized SPP model has particles which follow the equations of motion seen in Eqs. 1 and 2, however the parameters for each model are not constant in time. A particle is initialized with a random orientation and assigned an initial speed $v_0$ and rotational diffusion $D_r$ drawn from distributions $P(v_0) = \frac{|v_0|}{\sigma_v^2} e^{-\frac{(v_0-\mu_v)^2}{\sigma_v^2}}$ and $P(D_r) = \frac{1}{\sqrt{\pi\sigma_D^2}} e^{-\frac{(D_r-\mu_D)^2}{\sigma_D^2}}$. We evolve the particle position and orientation for a time $\tau$ drawn from $P(\tau) = \frac{1}{\tau_0} e^{\tau/\tau_0}$, where $\tau_0$ is the mean run time determined by experimental data. The particle then undergoes a tumbling event across one timestep where $D_r = 2\pi$, where the value of rotational diffusion is chosen to approximate an event where the orientation is completely randomized. After the tumble a new $v_0$ and $D_r$ are assigned until the next tumbling event. In contrast to a L\'{e}vy walk or standard SPP model, motility parameters are varied in time to replicate the variations and changes in a biological environment.

For both models, particle trajectories are constructed by numerically integrating the equations of motion using a simple Euler scheme with a timestep $dt = 0.1$. To compare these results to experimental data, we equate the simulation time unit to 4 minutes.

Finally, we note the VACF for experimental data shows a sharp dropoff across one frame due to errors in reconstructing the nuclei centers caused by imaging noise and fluctuations in dye intensity. To replicate this feature we incorporate positional noise into both models through small Gaussian fluctutations. After particle trajectories are constructed, each position is changed by a vector $\delta \vec{r} = dr \hat{\phi}$, where $dr$ is drawn from a Gaussian distribution of variable width $\Delta$ and the direction $\hat{\phi}$ is chosen randomly from the unit circle. This replicates experimental error in reconstructing cell positions, and allows our model trajectories to match the mouse fibroblast data.

\section{Results}

\subsection{Experimentally observed ensemble-averaged quantities are well fit by several existing models}

\begin{figure}
\centering
\includegraphics[width=0.45\textwidth]{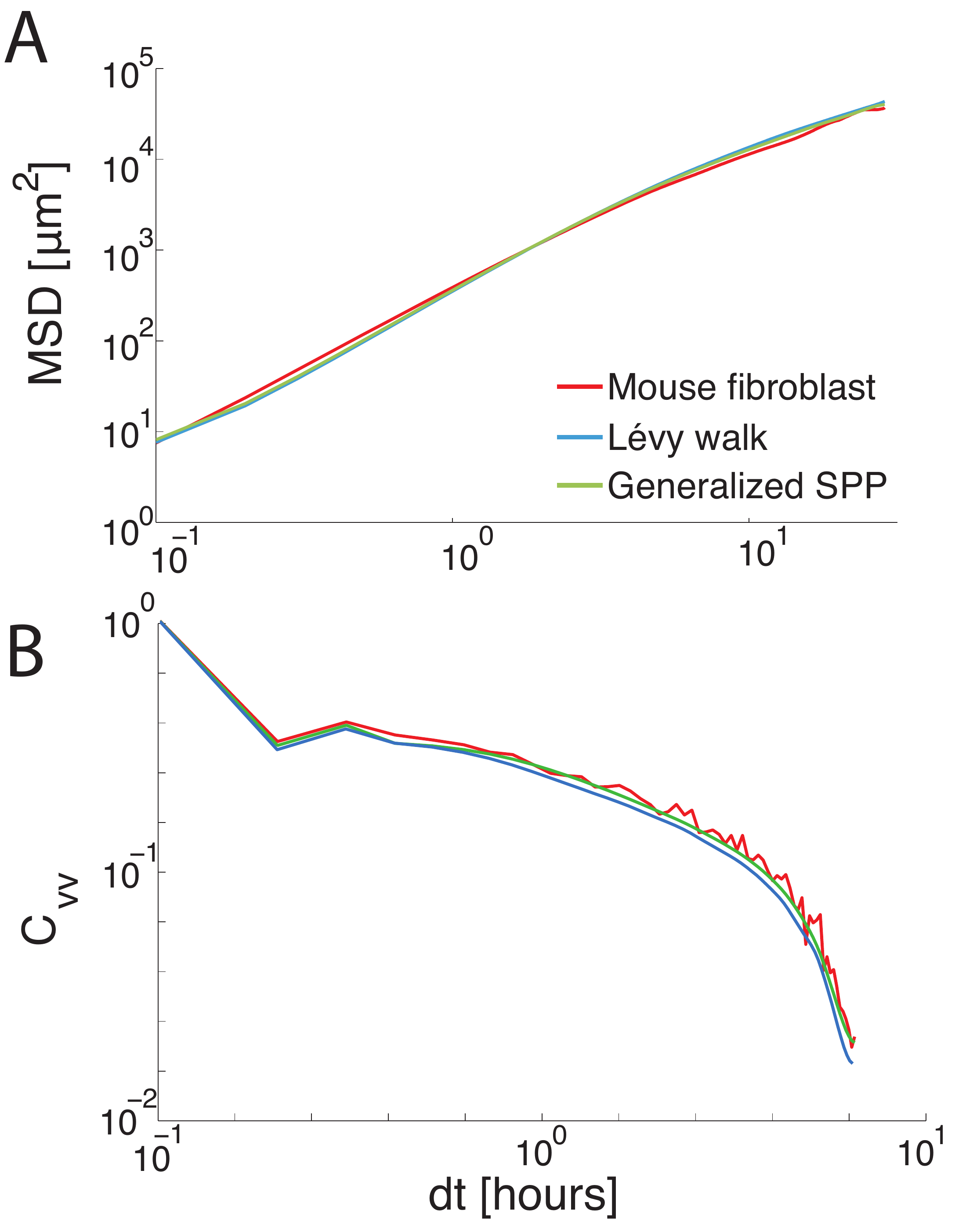}
\caption{\label{msd_compare}All of the proposed superdiffusive models are capable of capturing ensemble-averaged mouse fibroblast statistics.(A) Mean-squared displacements for mouse fibroblast cells, shown in red, as well as a L\'{e}vy walk and a generalized SPP model. Both models are able to match the mouse fibroblast MSD within the margin of error. (B) The velocity auto-correlation function $C_{vv}$ as a function of time $dt$.  There is a sharp decrease in the VACF across the first frame, due to error in reconstructing the nuclei centers-of-mass generated by imaging noise and fluctuations in dye intensity. At the largest timescales, each bin corresponds to fewer events and so error bars become large. In addition, adding positional error to simulation trajectories to match the initial dropoff in the VACF causes significant fluctuations at larger timescales.}
%less error from ensemble averaging. }
\end{figure}

Previous reports have compared models to experimental data using ensemble-averaged statistics such at the MSD and the VACF. Therefore, our first goal is to determine whether one of the existing models for explaining superdiffusive cell trajectories is a better fit to the experimental MSD and VACF data, shown by the red lines in Fig.~\ref{msd_compare}. 

For comparison, we simulate a L\'{e}vy walk model with dynamics given by Eqns 3 and 4, as well as a generalized SPP with no L\'{e}vy-walk behavior, described in more detail below. With the best-fit parameters, we find that both models match the data equally well.  As shown in Fig. \ref{msd_compare} (B), the velocity autocorrelation function exhibits a sharp decrease after the first frame window, due to errors that we make in reconstructing the nuclei center of mass caused by imaging noise and fluctuations in the dye intensity. Therefore, we add an additional term to the model that shifts the particle position by a Gaussian distributed variable with zero-mean and variance $\Delta^2$ to account for this effect. %\updated{The mouse fibroblast VACF is recovered for $\Delta = 1$, which is on the order of magnitude of particle speed $v_0$. This creates a caging effect where the mean normalized velocity across one timestep is dominated by these fluctuations and becomes less correlated than larger timesteps.}

\begin{figure*}
\centering
\includegraphics[width=0.8\textwidth]{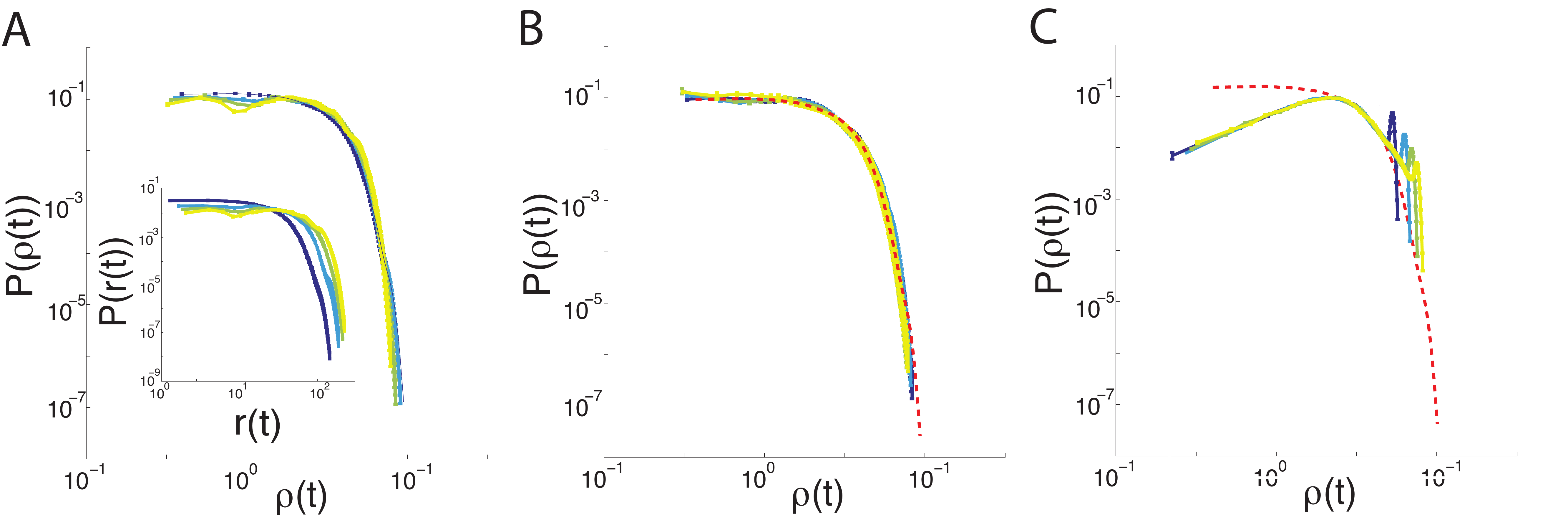}
\caption{\label{collapse} Displacement probability distribution $P(\rho)$, where $\rho(t)$ is the scaled displacement $\frac{r(t)}{t^{\gamma}}$, for the value of $\gamma$ that best collapses the data, for (A) Mouse fibroblast cells, (B) generalized SPP model and (C) L\'{e}vy walk, with colors representing 4 binned timescales from blue (small) to yellow (large). Mouse fibroblast $\tilde{P}(\rho)$ is shown as a dashed red line in (B,C) for comparison with each model, showing that only the generalized SPP model is consistent with the observed data. 
%(E) The $\chi^2$ goodness-of-fit approximation for each collapse shows the optimal value of scaling exponent $\gamma$ for each model. 
}
\end{figure*}

\begin{figure}
\centering
\includegraphics[width=0.4\textwidth]{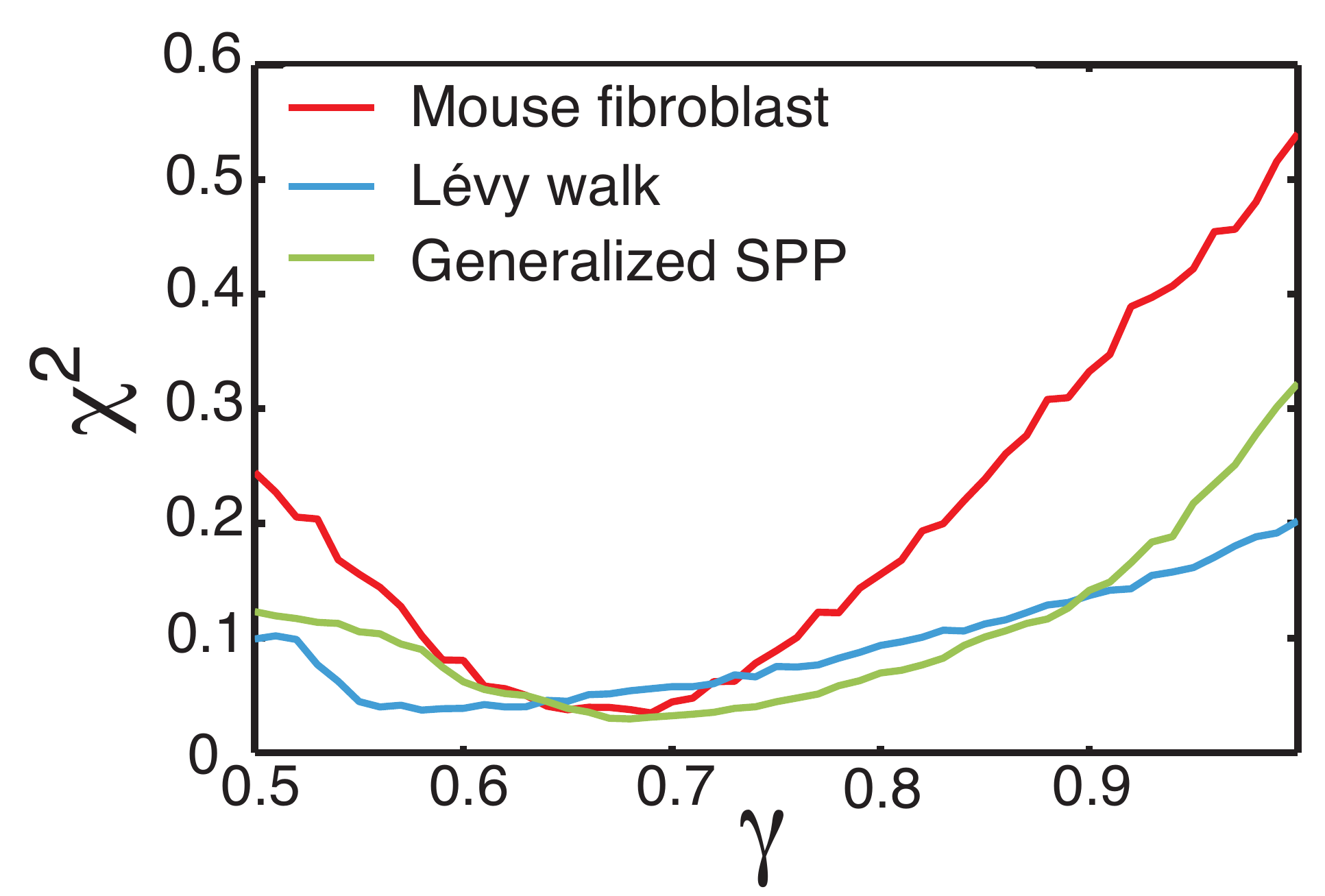}
\caption{\label{collapse2} Goodness-of-fit ($\chi^2$) as a function of scaling exponent $\gamma$. The value of $\gamma$ that best collapses each data set minimizes the $\chi^2$ goodness of fit between the $P(\rho(t))$ calculated at each timescale. Experimental data all collapse at a value of $0.5 < \gamma < 1$, consistent with a superdiffusive MSD.
%(E) The $\chi^2$ goodness-of-fit approximation for each collapse shows the optimal value of scaling exponent $\gamma$ for each model. 
}
\end{figure}

While the mean-squared displacement and velocity auto-correlation function are standard metrics for characterizing ensembles of trajectories, they may not be ideal for studying systems with superdiffusion. In an investigation of the L\'{e}vy walk properties of T-cells, Harris et al. study a quantity that reveals structures on shorter timescales: the probability for a cell to be at a displacement $r(t)$ at time $t$ \cite{Harris2012}. They suggest that Levy walks can be distinguished by collapsing these probability distributions with rescaled displacements $\rho(t) = \frac{r(t)}{t^{\gamma}}$, with $\gamma$ significantly larger than the value of $1/2$ expected for a persistent random walk. As seen in Fig. \ref{collapse}, we find that the mouse-fibroblast data does collapse, with the best fit exponent $\gamma = 0.69 \pm 0.02$ as shown in Fig. \ref{collapse2}. The best-fit Levy walk model collapses at $\gamma = 0.59 \pm 0.03$, which is above the value expected for a persistent random walk but still lower than $\gamma$ for mouse fibroblast cells. Importantly, the best-fit generalized SPP model also collapses at a similar value of $\gamma$, suggesting that such a collapse is not sufficient to uniquely identify L\'{e}vy walks as a mechanism for superdiffusivity.  

Moreover, the functional form of the displacement probability distribution $P(r(t))$ provides additional information. Fig. \ref{collapse} shows that $P(r(t))$ for the best-fit L\'{e}vy walk model has a significantly different functional form from mouse fibroblast trajectories at small displacements, due to ballistic runs over relatively large distances. In contrast, a non-L\'{e}vy version of the generalized SPP model matches the shape of mouse fibroblast $P(r(t))$ very well, providing an indication that a non-L\'{e}vy model might be better for describing mouse fibroblast data. 

\section{Numerical models are better constrained by single-cell trajectory data}

\begin{figure*}
\centering
\includegraphics[width=0.8\textwidth]{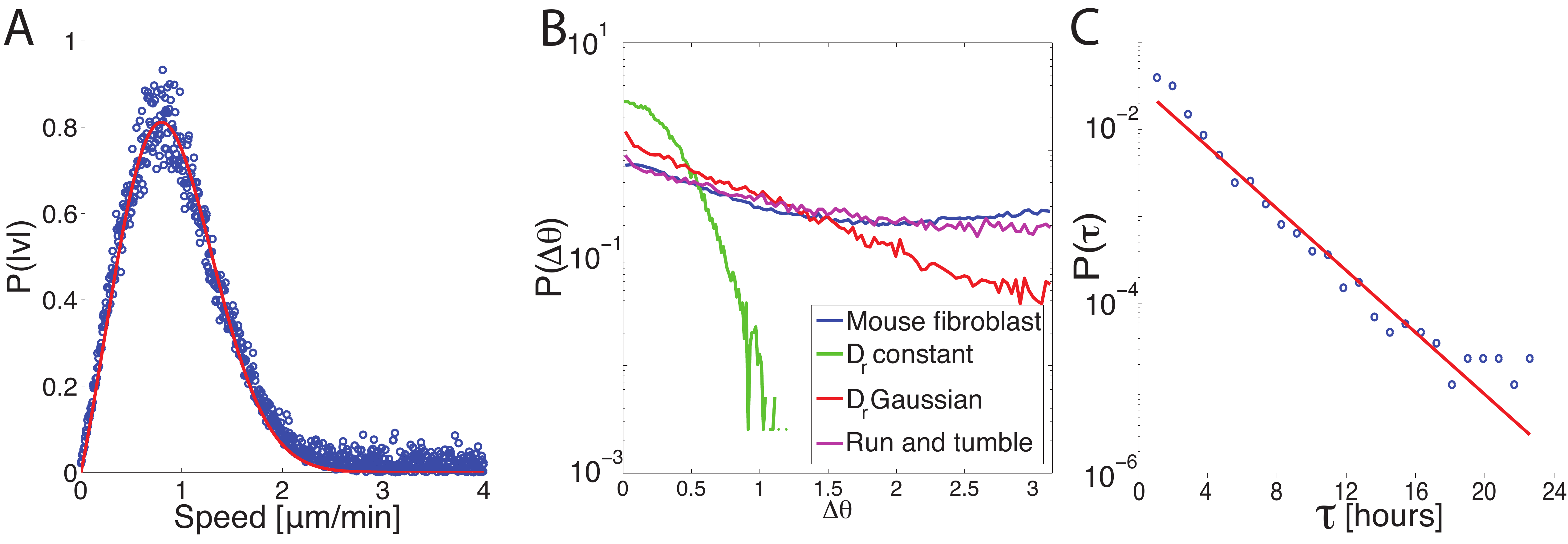}
\caption{\label{distributions}(A) Distribution of mouse fibroblast instantaneous speeds calculated from cell nuclei center-of-mass displacement between image capture events (blue). The red line is a fit to the form $P(v_0) = \frac{|v_0|}{\sigma_v^2} e^{-\frac{(v_0-\mu_v)^2}{\sigma_v^2}}$, which is the distribution of speeds expected for a Gaussian distribution of velocities. (B) Distribution of turning angles of mouse fibroblast trajectories (blue), SPP models with constant $D_r$ (green), Gaussian distributed $D_r$ (red), and a run-and-tumble model with Gaussian distributed $D_r$ during runs and exponentially distributed tumbling events. The distribution of rotational diffusion constants is the same in both heterogeneous cases to highlight the effect of incorporating tumbling events into the system. (C) Run-time distribution for mouse fibroblast cells (blue) is well fit by an exponential distribution (red). }
\end{figure*}

We next study single-cell trajectories. A generalized SPP model with arbitrary distributions for $P(v_0)$ and $P(D_r)$ has an infinite number of parameters that we could never hope to constrain. As a first step to simplifying our model we constrain functional form of these distributions using experimental data. As shown in Fig. \ref{distributions} (A), we first construct a distribution of cell speeds, determined from the magnitudes of the displacement of nuclei centers-of-mass between image capture events. Our experimental data is consistent with a Gaussian distribution of cell velocities, or equivalently, a distribution of cell speeds of the form $P(v_0) = \frac{|v_0|}{\sigma_v^2} e^{-\frac{(v_0-\mu_v)^2}{\sigma_v^2}}$, where $\mu_v$ and $\sigma_v$ are the mean and standard deviation, respectively. Therefore, we use this functional form in our generalized model. Next we estimate a distribution $P(D_r)$ of rotational diffusion constants ($D_r$) from the distribution of turning angles, shown in Fig. \ref{distributions} (B). Simple active Brownian systems with a single value of $D_r$ will generate a Gaussian distribution of turning angles \cite{Marchetti2013}. A Gaussian distribution of rotational noise broadens this distribution significantly. One can show the expected turning angle distribution in this case is a modified Bessel function of the second kind with an exponential tail, consistent with the numerical simulation data given by the red line in Fig. \ref{distributions} (B). We were unable to match the mouse fibroblast turning angle distribution, which is given by the blue line in Fig. \ref{distributions}(B) and has significant weight as the largest values of $\Delta\theta$, with any Gaussian function for the rotational noise. 

This suggests that mouse fibroblast cells may have a strongly bimodal distribution of rotational noises, further supported by intermittent run-and-tumble behavior seen in videos (Supplementary Movie 1.) We choose to capture this bimodal behavior with a noisy run-and-tumble model, where cells have a distribution $P(D_r) = \frac{1}{\sqrt{\pi\sigma_D^2}} e^{-\frac{(D_r-\mu_D)^2}{\sigma_D^2}}$ during runs, which are punctuated by tumbling events. We use the Canny algorithm described in the methods section to explicitly identify tumbling events, and the data points in Fig. \ref{distributions}(C) show the distribution of times between such events. The red line in \ref{distributions}(C) shows this is well-fit by an exponential distribution with with $\tau_0 \approx 1$ hour, and so in our model the distribution of run times $\tau$ is given by $P(\tau) = \frac{1}{\tau_0} e^{-\tau/\tau_0}$. We note that this is a strong indication that the mouse fibroblasts are not well-described by a L\'{e}vy walk model. The magenta line in Fig.\ref{distributions}(B) shows the distribution of turning angles for a noisy run-and-tumble model with the parameters identified above.

To confirm that the model parameters we have identified are robust, and to quantify their sensitivity, we vary model parameters around the microscopically determined values and quantify how much this changes their displacement probability distributions. Specifically, we use the linear regression goodness-of-fit parameter ($R^2$) between $P(r(t))$ for mouse fibroblast and generalized model trajectories to characterize each parameter configuration and identify a best-fit between our model and mouse fibroblast statistics \cite{corrCoeff}. Using this method we are able to capture the functional form of $P(r(t))$ very well, as shown in Fig. \ref{collapse}. It should be noted that incorporating a similar distribution of speeds into a L\'{e}vy walk model would improve the fit seen in Fig. \ref{collapse}C. However the distribution of run times would remain a power law and not match the distribution of mouse fibroblast run times shown in Fig. \ref{distributions}C, which is fit well by an exponential.

\begin{figure*}
\centering
\includegraphics[width=0.8\textwidth]{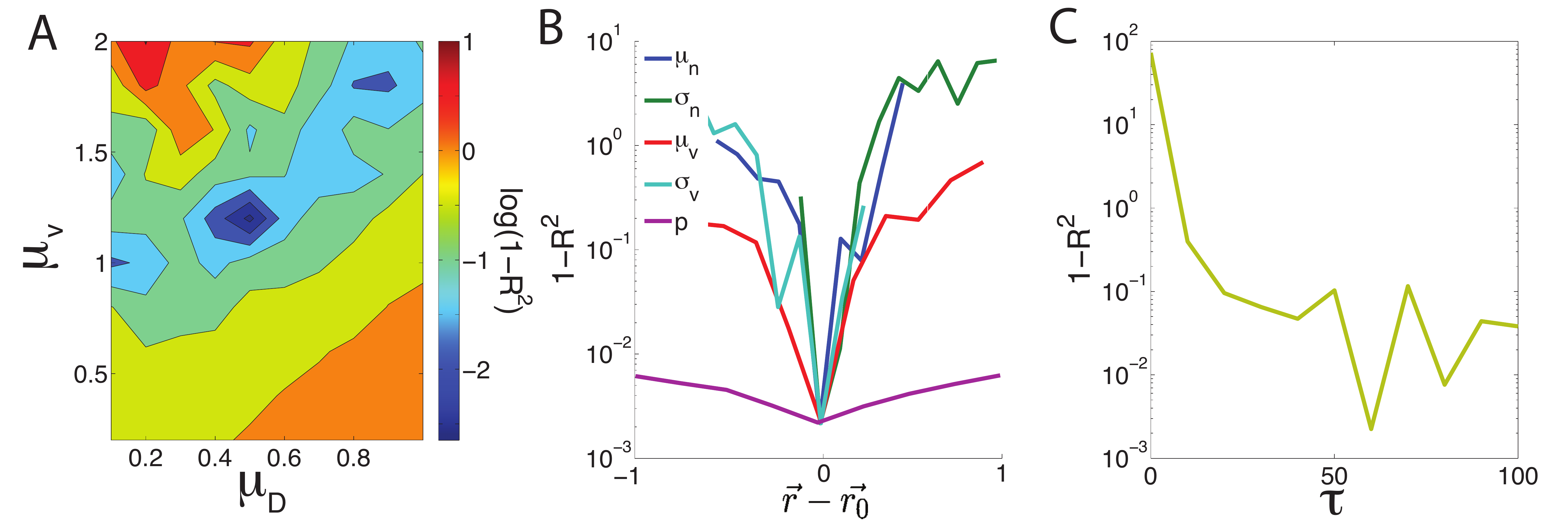}
\caption{\label{goodbad} Sensitivity analysis examining the goodness of fit of the generalized SPP model to displacement distributions in mouse fibroblasts $(1-R^2)$ as a function of model parameters. (A) Contour plot of $log(1-R^2)$ illustrates the experimental data tightly constrains a linear combination of the mean velocity $\mu_v$ and mean noise $\mu_D$. (B,C) The goodness of fit as a function of each model parameter while all others are held fixed. (B) $\vec{r_0}$ is the optimal coordinate in parameter space and $\vec{r}- \vec{r_0}$ is the distance of each parameter from its optimal value. (C) A value of $\tau$ smaller than $\approx 10$ is inconsistent with experimental data, but data does not discriminate between larger values of $\tau$. }
\end{figure*}

Happily, the configuration of parameters that best matches the macroscopic P(r(t)), located at  $\mu_{D} = 0.09, \sigma_{D} = 0.002, \mu_{v} = 1, \sigma_{v} = 0.8, p = 0, \tau_0= 10$, is very similar to those identified from microscopic statistics, indicating that the model is consistent with experimental results.  A construction of the dynamical matrix around this minima and subsequent analysis of local eigenvectors indicates that our system is most sensitive to perturbations in the mean velocity and mean rotational noise as shown in Fig. \ref{goodbad}A, and relatively insensitive to correlations between $D_r$ and $v_0$ parameterized by $p$ (Fig~\ref{goodbad}B) as well as mean run time $\tau_0$ (Fig~\ref{goodbad}C).

\section{Discussion} 
Both L\'{e}vy walks and heterogeneous SPP models are capable of generating superdiffusive trajectories. Previous studies have focused on one model or the other in order to identify possible mechanisms for superdiffusive cell trajectories.

We show that while both types of model are equally capable of matching the large-scale ensemble averaged statistics of mouse fibroblast cells, an analysis of single cell trajectories demonstrates that L\'{e}vy walks are not consistent with this data set, despite a very good scaling collapse of the probability displacement distribution with scaling exponent $\gamma > 1/2$. Instead, a careful analysis of turning angle distributions suggests thse mouse fibroblasts exhibit heterogeneous speeds, with noisy run-and-tumble behavior. 

Because superdiffusive cells are able to cover distance faster than diffusive counterparts, it would be useful to adapt the tools developed here to study many more cell types.  For example, directed cell motion is known to be a signature of invasiveness in cancer cell lines~\cite{Driscoll2012}, and it would be interesting to know if these cell types are  altering the mechanisms or timescales for superdiffusion as they become more malignant. To that end, we have created a MATLAB software package for deploying these analyses on generic data sets~\cite{ManningGroup}, which can be used to quantify superdiffusive dynamics and distinguish between different mechanism behavior in cells and active matter. 

A natural extension of our current work is interacting SPP models. While a non-interacting model can approximate our mouse fibroblast data where cells are not in constant contact, higher density cell populations, and confluent tissues will require models with steric cell-cell interactions. The effect of super-diffusion, whether generated by a L\'{e}vy walk or heterogeneity based model, could potentially alter results obtained with standard SPP models.

For example, recent work suggests that groups of cells \cite{Bi2016} and packings of SPPs undergo jamming transitions ~\cite{Henkes2011,Fily2012,Berthier2014}. Could the addition of superdiffusive dynamics have an effect on these types of transitions? Persistent motility can alter the jamming transition -- higher speeds and more persistent trajectories allows particles to explore areas of the energy landscape that were previously inaccessible \cite{Berthier2014}. Similar effects are seen in shape-based models for confluent tissues~\cite{Bi2016}. The inclusion of both run-and-tumble dynamics as well as varying persistence length through broadly distributed rotational diffusion coefficients in a generalized SPP model could offer an interesting mechanism for tuning jamming. 

Another emergent feature of self-propelled particle models is motility induced phase separation (MIPS). Persistently moving particles create an inward oriented boundary layer that cage interior particles into a solid phase, while other cells are in a lower density gas phase outside of this boundary \cite{Fily2012,Cates2015} and this effect has recently been implicated in generating colony formation in bacteria \cite{Patch2017}. MIPS relies on persistence length to generate this behavior. Our generalized SPP model could reinforce this effect due to relatively persistent run phases, destroy the effect due to tumbling, or perhaps alter the nature of the transition due to enhanced fluctuations, and this is an interesting direction for future study. 

Density also plays a critical role in cell interactions. Many cell types exhibit contact inhibition of locomotion (CIL), where contact with another cell will either halt their motion of cause them to immediately recoil and begin moving moving in the opposite direction. It is possible that the tumbling events we see in mouse fibroblast cells are CIL events.  There could also be additional interactions such as alignment between neighbors or between cells and the underlying substrate to generate flocking~\cite{Vicsek1995}. It would be interesting to explore the effect of alignment in a generalized SPP model, to see if heterogeneity causes any significant differences in the flocking transition.

Another benefit of simple SPP models is that they can be relatively easily coarse-grained to predict large scale features of a tissue or colony \cite{Marchetti2013}. We have shown that a generalized SPP model is more consistent with superdiffusive mouse fibroblast cell trajectories than a L\'{e}vy walk, opening the door to a hydrodynamic coarse-graining approach for this system. 

% In a L\'{e}vy walk, the run-time distribution is heavy tailed, a signature of large scale effects in a system. A coarse graining of L\'{e}vy walk equations of motion may be more suited to investigate complex systems with possible long-range interactions such as confluent tissues. 

%Generation of super-diffusive trajectories using computational methods is well understood, however the biological origin of a L\'{e}vy distributed timescale is more elusive. Moving forward, more rigorous analysis of cell trajectory data using displacement probability distributions is likely to yield more useful information about biological systems.  

Until now, mouse fibroblasts have not been highlighted as a system with run-and-tumble behavior and therefore the biomolecular mechanisms responsible for this behavior are unknown. To begin to investigate this question, it would be useful to correlate tumbling events with the dynamics of sub-cellular features such as spatio-temporal distributions of focal adhesions~\cite{Dennis2011}, Golgi bodies~\cite{Deakin2014}, or actin waves~\cite{Driscoll2012}. This would help us to understand which signaling networks and components of motility machinery are involved generating tumbling behavior or broad distributions of rotational diffusion. Furthermore, it might be useful to study such behavior on structured or controllable substrates \updated{\cite{Riveline}}, to tease apart the influence of environment vs. internal circuitry on controlling these timescales.

\begin{acknowledgments}
We acknowledge helpful discussions with A. A. Middleton. All authors acknowledge support from NSF-BMMB-1334493. MLM and GP were supported by NSF-DMR-1352184, the Research Corporation and the Alfred P. Sloan Foundation, and GP and MB were supported by the IGERT program (NSF-DGE1068780). Funding for VZ provided by Russian Science Foundation Grant No. 16-12-10496.
\end{acknowledgments}

\end{article}

\newpage
\onecolumn
\section{Supplemental Material}
\subsection{Supplemental method 1}

C3H10T1/2 mouse fibroblast cells (ATCC) were cultured in Basal Medium Eagle complete growth medium supplemented with 10\% fetal bovine serum (v/v), 1\% penicillin/streptomycin (v/v), and 1\% GlutaMax (v/v). Cells were expanded in a 37$^{\circ}$C humidified incubator with regulated 5\% CO$_2$ and passaged at 80\% confluence using 0.25\% Trypsin EDTA. For time-lapse experiments, cells were restricted to passage numbers 12-18.
                        
            Prior to cell seeding, substrate samples were soaked in room temperature BME medium for 6 hrs to promote FBS protein adsorption to the material surface. Each sample was transferred into an individual well in a 48-well plate and cells were solution seeded (500 $\mu$L/well) at a density of 4000 cells/cm$^2$. Cell samples were then incubated at 30$^{\circ}$C for 16 hours to establish equilibrium prior to time-lapse image set-up.
            
            Hoechst nuclear stain was prepared at a concentration of 0.01 $\mu$g/mL in BME complete medium (30$^{\circ}$C). 800 $\mu$L of the staining solution was added to each well of a 4-well LabTek borosilicate chamber slide (Fisher Scientific) and incubated at 30$^{\circ}$C for 1 hr. Samples were then inverted and weighed down with sterilized glass slide inserts, cut to fit into the chamber wells. The chamber slide was then transferred to a live cell stage incubator (INC-2000, 20/20 Technology, Inc.) and cells were imaged using a Leica DMI 6000B inverted microscope. The live cell stage incubator was equilibrated at 30$^{\circ}$C with constant 5\% CO$_2$. One image per position of interest was captured every five minutes in phase, A4 (excitation/emission peak of 360/470 nm), and N3 (excitation/emission peak of 546/600 nm) using 50 ms, 100 ms, and 50 ms exposure times respectively on an Andor Luca R camera with a 10x/0.63 NA objective. Samples were imaged in succession for 4 hrs at 30$^{\circ}$C, followed by 20 hrs at 37$^{\circ}$C. The resultant motility image stacks were analyzed using the ACT\textit{IV}E image analysis package to track nuclei centers-of-mass~\cite{Baker2014}. 

\subsection{Supplemental method 2}

Cell motility was characterized using statistical analysis of cell nuclei trajectories. In general, the MSD for a collection of particle tracks is calculated as a function of time window ($\Delta t$):
\begin{equation}
MSD(\Delta t) = \frac{1}{N} \sum_{i = 1}^{N} \sum_{t = 1}^{T-\Delta t} [r_i(t+\Delta t) - r_i(t)]^2,
\end{equation}
where $N$ is the total number of particles (cells), $T$ is the total number of timesteps (frames), and $r_i(t)$ is the position of particle $i$ at time $t$. An example of the MSD extracted from tracked mouse fibroblasts is shown in Fig. \ref{msd_compare} (A).  As mentioned previously, the MSD can often be fit to the following functional form, 
\begin{equation}
MSD(\Delta t) \propto {\Delta t}^{\alpha}.
\end{equation}
We extract the exponent $\alpha$ from data using a linear fit of $log_{10}(MSD)$ vs. $log_{10}(\Delta t)$. The dependence of fit on timescale $\Delta t$ is extracted using a linear fit between timescales $\Delta t$ and $\Delta t + T$ with $T = 100$. Altering the fitting window $T$ did not significantly alter our results. Throughout the remainder of this manuscript we will denote the slope of the log-log MSD as $\alpha$. 

We calculate the standard directional velocity auto-correlation function:
\begin{equation}
C_{vv} = \left<\hat{v}(t_o)\hat{v}(t_o+dt)\right> = \frac{\vec{v}(t_o)}{|\vec{v}(t_o)|}\frac{\vec{v}(t_o+dt)}{|\vec{v}(t_o+dt)|},
\end{equation}
where the velocities, $\vec{v}(t)$, are the instantaneous displacements between two sequential frames and the brackets indicate averaging over the ensemble and initial times $t_o$. An example is shown in Fig. \ref{msd_compare} (B).

Unscaled displacement probability distributions were calculated by first constructing a cumulative distribution function (CDF) using the MATLAB function ksdensity. The probability distribution function is constructed as the numerical derivative of the CDF. This process was then repeated with scaled displacements $\rho(t) = \frac{r(t)}{t^{\gamma}}$ with the best collapse determined by minimizing the sum over the squared difference between each probability distribution and normalizing by the number of elements, sampling values of $0.4 < \gamma < 1$. Utilizing a bootstrap method of error estimation, the distributions from which the CDFs were constructed were randomly sampled with replacement to generate a new data set and the corresponding probability distribution. This process was iterated 100 times in order to estimate the variance of each bin. 

The Canny edge detection algorithm was used to calculate turning angle distributions \cite{Canny}. A key parameter in the Canny algorithm is the threshold for distinguishing a true turn from noise. This is calculated using Otsu's method on individual cell trajectories \cite{Otsu}, which uses a test angle magnitude, $k$, to divide the turning angles for a cell trajectory into two classes, runs and tumbles. The variance of each of these classes is then calculated as a function of $k$ and minimized to determine the optimal threshold value for the Canny edge detection algorithm. We define the run time as the time between edge detection in orientation space. Supplementary movie 2 demonstrates that our edge detection algorithm successfully identifies tumbling events during cell migration. An example trajectory with highlighted tumbling events is shown in Fig. \ref{mf_example}(B).

\end{document}